\def\lm{{\mathop{\rm lm}\nolimits}}
\def\init{{\mathop{\rm init}\nolimits}}
\def\prem{{\mathop{\rm prem}\nolimits}}
\def\rem{{\mathop{\rm rem}\nolimits}}
\def\pprecond{{\mathop{\rm pprecond}\nolimits}}
\def\sprem{{\mathop{\rm sprem}\nolimits}}
\def\rem{{\mathop{\rm rem}\nolimits}}
\def\sprecond{{\mathop{\rm sprecond}\nolimits}}
\def\card{{\tt card}}
\def\sotd{{\tt sotd}}

\def\td{{\tt td}}
\def\C{{\bf C}}

\def\R{{\bf R}}

\def\Gr{$=_G$}
\def\Greduce{${\buildrel*\over\rightarrow}^G$}
\def\GrC{$=_{GC}$}
\def\GrR{$=_{GR}$}
\def\Tr{$=_\Delta$}
\def\Col{$<_{\rm Col}$}
\def\QECol{\hbox{$\not\exists_{\rm CH}$}} 
\def\ECol{$=_{\rm Col}$}
\def\MMR{$<_{\Delta\R}$}
\def\MMC{$\ne_{\Delta\C}$}

\def\tableA{\centerline{
\begin{tabular}{|l|rr|rrr|rr|rrr|}
\hline
                  & \multicolumn{2}{|c|}{\QECol}     & \multicolumn{3}{|c|}{\Gr/\QECol}   & \multicolumn{2}{|c|}{\QECol/{\tt full-cad}}      & \multicolumn{3}{|c|}{\Gr/\QECol/{\tt full-cad}}                \\
  & \textbf{Time} & \textbf{Cells}   & \multicolumn{2}{c}{\textbf{Time}} &  \textbf{Cells} & \textbf{Time} & \textbf{Cells}   & \multicolumn{2}{c}{\textbf{Time}} &  \textbf{Cells} \\\hline
I A    & 190 & 503 & 22+72= & 94 & 23             & 188   & 503   & 22+73=  & 95    & 51    \\
I B    & 199 & 369 & 21+74= & 95 & 17             & 191   & 369   & 21+75=  & 96    & 33    \\
R A          & 85 & 1 & 24+73= & 97 & 1                 & 86    & 1     & 24+71=  & 95    & 1     \\
R B          & 129 & 1 & 24+72= & 96 & 1                & 125   & 1     & 24+72=  & 96    & 1     \\ 
E A         & 297 & 621 & 25+134= & 159 & 621          & 576   & 11139 & 25+394= & 419   & 11139 \\
E B         & Error & ?     & 50+?=   & Error & ?      & Error & ?     & 50+?=   & Error & ?     \\
S A      & 89 & 153 & 22+72=& 94 & 63               & 199   & 349   & 22+185= & 207   & 625   \\
S B      & 113 & 375 & 23+75= & 98 & 41             & 228   & 1063  & 23+180= & 203   & 237   \\
C A       & 133 & 19 & 42+?= & Error & ?             & 235   & 19    & 42+?=   & Error & ?     \\
C B       & Error & ?     & 132+?=  & Error & ?      & Error & ?     & 132+?=  & Error & ?     \\\hline
\end{tabular}}
}

\def\tableB{\centerline{
\begin{tabular}{|l|r|r|r|r|r|r|}
\hline
                 & \MMC & \MMR & & \Gr/\MMC & \Gr/\MMR  & \\
 & \textbf{Time} & \textbf{Time} & \textbf{Ratio} & \textbf{Time} & \textbf{Time} & \textbf{Ratio} \\\hline
Intersection A   & 5691          & 29426         & 4.17           & 1168          & 2470          & 1.11           \\
Intersection B   & 5584          & 36262         & 5.49           & 886           & 1482          & 0.67           \\
Random A         & 4614          & 17355         & 2.76           & 310           & 570           & 0.84           \\
Random B         & 67343         & 356670        & 4.30           & 318           & 470           & 0.48           \\
Ellipse A*       & 85425         & 262623        & 2.07           & 27916         & 62496         & 1.24           \\
Ellipse B*       & 441245        & $>1000s$      & -              & $>1000s$      & $>1000s$      & -              \\
Solotareff A*    & 6666          & 16014         & 1.40           & 1760          & 2025          & 0.15           \\
Solotareff B*    & 9536          & 43439         & 3.56           & 1404          & 1647          & 0.17           \\
Collision A*     & 41085         & 216028        & 4.26           & $>1000s$      & $>1000s$      & -              \\
Collision B*     & $>1000s$      & $>1000s$      & -              & $>1000s$      & $>1000s$      & -              \\\hline
\end{tabular}}
}

\def\tableC{\centerline{
\begin{tabular}{|l|rr|rrr|rr|}
\hline
                        & \multicolumn{2}{|c|}{\MMR} & \multicolumn{3}{|c|}{\Gr/\MMR}& \multicolumn{2}{|c|}{Ratio}\\
  & \textbf{Time}      & \textbf{Cells}           & \multicolumn{2}{|c}{\textbf{Time}}            & \textbf{Cells}            & \textbf{Time} & \textbf{Cells} \\\hline
Cyclic--3               & 3136               & 381                      & 20 + 245 =&265               & 21                        & 11.83         & 18.14          \\ 
Cyclic--4               & $>1000s$           & ?                        & 64 + 5813 =&5877             & 621                       & ?             & ?              \\
2                       & 2249               & 895                      & 22 + 1845 =&1867             & 579                       & 1.20          & 1.55           \\ 
4                       & 3225               & 421                      & 24 + 19738 =&19762           & 1481                      & 0.16          & 0.28           \\
6                       & 363                & 41                       & 20 + 918 =&938               & 89                        & 0.39          & 0.46           \\ 
7                       & 3667               & 895                      & 26 + 6537 =&6563             & 1211                      & 0.56          & 0.74           \\ 
8                       & 3216               & 365                      & 21 + 174 =&195               & 51                        & 16.49         & 7.16           \\ 
13                      & 14342              & 4949                     & 18 + 220 =&238               & 81                        & 60.26         & 61.10          \\
14                      & 334860             & 27551                    & 21 + 971 =&992               & 423                       & 337.56        & 65.13          \\\hline
\end{tabular}}
}

\def\tableD{\centerline{
\begin{tabular}{|l|rr|rr|rr|rr|rr|}
\hline
& \multicolumn{2}{|c|}{\MMR} & \multicolumn{2}{|c|}{\Gr/\MMR} & \multicolumn{2}{|c|}{\Gr/$\rightarrow^G_y$/\MMR} & \multicolumn{2}{|c|}{\Gr/$\rightarrow^G_x$/\MMR} & \multicolumn{2}{|c|}{\Gr/$\stackrel{\ast}{\rightarrow}^G$/\MMR}\\
 & \textbf{Time} & \textbf{Cells} & \textbf{Time} & \textbf{Cells} & \textbf{Time} & \textbf{Cells} & \textbf{Time} & \textbf{Cells} & \textbf{Time} & \textbf{Cells} \\\hline
${S}_1,{S}_2,{C}$  & 9830          & 1073           & 1057          & 267            & 394           & 91             & 528           & 183            & 298           & 99             \\
${S}_2,{S}_3,{C}$  & 187048        & 12097          & 5880          & 1299           & 3171          & 627            & 2149          & 517            & 506           & 213            \\
${S}_3,{S}_4,{C}$  & 247458        & 11957          & 8164          & 1359           & 9177          & 1123           & 5476          & 881            & 590           & 213            \\\hline
\end{tabular}}
}

\def\tableE{\centerline{
\begin{tabular}{|l|rr|rrr|rrr|}
\hline
 & \multicolumn{2}{|c|}{\Col} & \multicolumn{3}{|c|}{\Gr/\Col} & \multicolumn{3}{|c|}{\Gr/$\stackrel{\ast}{\rightarrow}^G$/\Col}  \\
 & \textbf{Time} & \textbf{Cells} & \multicolumn{2}{|c}{\textbf{Time}} & \textbf{Cells} & \multicolumn{2}{|c}{\textbf{Time}}  & \textbf{Cells} \\\hline
${S}_1,{S}_2,{C}$  & 30            & 1073           & 23 + 8  = & 31       & 267            & 24 + 4  = & 28        & 99             \\
${S}_2,{S}_3,{C}$  & 763           & 12097          & 27 + 36 = & 63       & 1299           & 28 + 13 = & 41        & 213            \\
${S}_3,{S}_4,{C}$  & 1760          & 11957          & 28 + 37 = & 65       & 1359           & 29 + 14 = & 43        & 213            \\\hline
\end{tabular}}
}

\def\tableFa{\centerline{
\begin{tabular}{|l|l|rr|rr|rr|}
\hline
\multicolumn{2}{|c|}{} & \multicolumn{2}{|c|}{\MMR} & \multicolumn{2}{|c|}{\Gr/\MMR} & \multicolumn{2}{|c|}{\Gr/\Greduce/\MMR}\\
\multicolumn{2}{|c|}{} & \textbf{Time} & \textbf{Cells}  & \textbf{Time} & \textbf{Cells}  & \textbf{Time} & \textbf{Cells}  \\\hline
${S}_1,{S}_2,{C}$ & C & 8654   & 1073  & 905    & 267  & 270    & 99    \\
                  & R &        &       & 902    & 267  & 453    & 183   \\\hline
${S}_2,{S}_3,{C}$ & C & 189202 & 12097 & 5911   & 1299 & 499    & 213   \\
                  & R &        &       & 18941  & 2639 & 5307   & 859   \\\hline
${S}_3,{S}_4,{C}$ & C & 248340 & 11957 & 8159   & 1359 & 580    & 213   \\
                  & R &        &       & 160171 & 9091 & 196714 & 11203 \\\hline
\end{tabular}}
}

\def\tableFb{\centerline{
\begin{tabular}{|l|rrr|rrr|rrr|}
\hline
 & \multicolumn{3}{|c|}{\MMR} & \multicolumn{3}{|c|}{\Gr/\MMR} & \multicolumn{3}{|c|}{\Gr/\Greduce/\MMR} \\
 & degrees & \textbf{Time} & \textbf{Cells} & degrees & \textbf{Time} & \textbf{Cells} & degrees & \textbf{Time} & \textbf{Cells} \\\hline
${S}_1,{S}_2,{C}$ & 6 / 18 & 8654   & 1073  & 5 / 9  & 905    & 267  & 5 / 7  & 270    & 99    \\
${S}_2,{S}_3,{C}$ & 6 / 19 & 189202 & 12097 & 5 / 11 & 5911   & 1299 & 5 / 10 & 499    & 213   \\
${S}_3,{S}_4,{C}$ & 6 / 21 & 248340 & 11957 & 5 / 15 & 8159   & 1359 & 5 / 15 & 580    & 213   \\\hline
\end{tabular}}
}

\def\tableGa{\centerline{
\begin{tabular}{|l|l|rr|rr|}
\hline
\multicolumn{2}{|c|}{} & \multicolumn{2}{|c|}{\MMR} & \multicolumn{2}{|c|}{\Gr/\MMR} \\
\multicolumn{2}{|c|}{} & \textbf{Time} & \textbf{Cells} & \textbf{Time} & \textbf{Cells} \\\hline
Intersection A & C & 29426    & 3763  & 2470     & 273   \\
               & R &          &       & $>1000s$ & ?     \\\hline
Intersection B & C & 36262    & 2795  & 1482     & 189   \\
               & R &          &       & $>1000s$ & ?     \\\hline
Random A       & C & 17355    & 1219  & 570      & 165   \\
               & R &          &       & $>1000s$ & ?     \\\hline
Random B       & C & 356670   & 7119  & 470      & 141   \\
               & R &          &       & $>1000s$ & ?     \\\hline
Ellipse A*     & C & 262623   & 28557 & 62496    & 14439 \\
               & R &          &       & 271726   & 29939 \\\hline
Ellipse B*     & C & $>1000s$ & ?     & $>1000s$ & ?     \\
               & R &          &       & $>1000s$ & ?     \\\hline
Solotareff A*  & C & 16014    & 1751  & 2025     & 297   \\
               & R &          &       & $>1000s$ & ?     \\\hline
Solotareff B*  & C & 43439    & 6091  & 1647     & 243   \\
               & R &          &       & $>1000s$ & ?     \\\hline
Collision A*   & C & 216028   & 7895  & $>1000s$ & ?     \\
               & R &          &       & $>1000s$ & ?     \\\hline
Collision B*   & C & $>1000s$ & ?     & $>1000s$ & ?     \\
               & R &          &       & $>1000s$ & ?     \\\hline
\end{tabular}}
}

\def\tableGb{\centerline{
\begin{tabular}{|l|rrr|rrr|}
\hline
 & \multicolumn{3}{|c|}{\MMR} & \multicolumn{3}{|c|}{\Gr/\MMR} \\
  & degrees & \textbf{Time} & \textbf{Cells} & degrees & \textbf{Time} & \textbf{Cells} \\\hline
Intersection A & 6  / 14 & 29426    & 3763  & 17 / 50  & 2470     & 273   \\
Intersection B & 6  / 14 & 36262    & 2795  & 15 / 41  & 1482     & 189   \\
Random A       & 9  / 16 & 17355    & 1219  & 19 / 68  & 570      & 165   \\
Random B       & 9  / 16 & 356670   & 7119  & 19 / 73  & 470      & 141   \\
Ellipse A*     & 6  / 24 & 262623   & 28557 & 6  / 26  & 62496    & 14439 \\
Ellipse B*     & 6  / 24 & $>1000s$ & ?     & 25 / 253 & $>1000s$ & ?     \\
Solotareff A*  & 10 / 25 & 16014    & 1751  & 10 / 28  & 2025     & 297   \\
Solotareff B*  & 10 / 25 & 43439    & 6091  & 21 / 69  & 1647     & 243   \\
Collision A*   & 6  / 23 & 216028   & 7895  & 27 / 251 & $>1000s$ & ?     \\
Collision B*   & 6  / 23 & $>1000s$ & ?     & 36 / 875 & $>1000s$ & ?     \\\hline
\end{tabular}}
}

\def\tableAb{\centerline{
\begin{tabular}{|l|rr|rrr|rr|rr|}
\hline
                  & \multicolumn{2}{|c|}{\ECol} & \multicolumn{3}{|c|}{\Gr/\ECol} & \multicolumn{2}{|c|}{\MMR} & \multicolumn{2}{|c|}{\Gr/\MMR} \\
  & \textbf{Time} & \textbf{Cells}   & \multicolumn{2}{c}{\textbf{Time}} &  \textbf{Cells}           & \textbf{Time} & \textbf{Cells}         & \textbf{Time} & \textbf{Cells}                 \\\hline
I A    & 236   & 3723  & 22+77=  & 99    & 273   & 29426    & 3763   & 2470     & 273   \\
I B    & 212   & 3001  & 21+76=  & 97    & 189   & 36262    & 2795   & 1482     & 189   \\
R A          & 150   & 2101  & 24+86=  & 110   & 105   & 17355    & 1267   & 570      & 165   \\
R B          & 21091 & 7119  & 24+80=  & 104   & 141   & 356670   & 7119   & 470      & 141   \\ 
E A*        & 7390  & 114541& 25+3189=& 3214  & 53559 & 262623   & 28557  & 62496    & 14439 \\
E B*        & Error & ?     & 50+?=   & Error & ?     & $>1000s$ & ?      & $>1000s$ & ?     \\
S A*     & 115   & 1751  & 22+82=  & 104   & 297   & 16014    & 1751   & 2025     & 297   \\
S B*     & 253   & 6091  & 23+82=  & 105   & 243   & 43439    & 6091   & 1647     & 243   \\
C A*      & 820   & 8387  & 42+?=   & Error & ?     & 216028   & 7895   & $>1000s$ & ?     \\
C B*      & Error & ?     & 132+?=  & Error & ?     & $>1000s$ & ?      & $>1000s$ & ?     \\\hline
\end{tabular}}
}

\def\tableH{\centerline{
\begin{tabular}{|l|rrr|rrr|rrr|}
\hline
\multicolumn{1}{|c|}{} & \multicolumn{3}{|c|}{\MMR} & \multicolumn{3}{|c|}{\Gr/\MMR} & \multicolumn{3}{|c|}{\Gr/\Greduce/\MMR}\\
\multicolumn{1}{|c|}{}  & \texttt{TNoI} & \textbf{Time} & \textbf{Cells}  & \texttt{TNoI} & \textbf{Time} & \textbf{Cells} & \texttt{TNoI}  & \textbf{Time} & \textbf{Cells}  \\\hline
${S}_1,{S}_2,{C}$     & 8 & 8654   & 1073  & 5 & 905    & 267  & 4 & 270    & 99    \\
${S}_2,{S}_3,{C}$     & 8 & 189202 & 12097 & 6 & 5911   & 1299 & 6 & 499    & 213   \\
${S}_3,{S}_4,{C}$     & 8 & 248340 & 11957 & 7 & 8159   & 1359 & 7 & 580    & 213   \\\hline
\end{tabular}}
}
\def\tableI{\centerline{
\begin{tabular}{|l|rrr|rrr|}
\hline
\multicolumn{1}{|c|}{} & \multicolumn{3}{|c|}{\MMR} & \multicolumn{3}{|c|}{\Gr/\MMR} \\
\multicolumn{1}{|c|}{}  & \texttt{TNoI} & \textbf{Time} & \textbf{Cells} & \texttt{TNoI} & \textbf{Time} & \textbf{Cells} \\\hline
Intersection A     & 8 & 29426    & 3763  & 7  & 2470     & 273   \\
Intersection B     & 8 & 36262    & 2795  & 7  & 1482     & 189   \\
Random A           & 9 & 17355    & 1219  & 5  & 570      & 165   \\
Random B           & 9 & 356670   & 7119  & 5  & 471      & 141   \\
Ellipse A*         & 7 & 262623   & 28557 & 6  & 62496    & 14439 \\
Ellipse B*         & 7 & $>1000s$ & ?     & 21 & $>1000s$ & ?     \\
Solotareff A*      & 9 & 16014    & 1751  & 8  & 2025     & 297   \\
Solotareff B*      & 9 & 43439    & 6091  & 7  & 1647     & 243   \\
Collision A*       & 7 & 216028   & 7895  & 18 & $>1000s$ & ?     \\
Collision B*       & 7 & $>1000s$ & ?     & 22 & $>1000s$ & ?     \\\hline
\end{tabular}}
}

\def\tableJ{\centerline{
\begin{tabular}{|l|rrr|rrrr|rr|}
\hline
                        & \multicolumn{3}{|c|}{\MMR} & \multicolumn{4}{|c|}{\Gr/\MMR}\\
 & \texttt{TNoI} &\textbf{Time}       & \textbf{Cells}   & \texttt{TNoI}  & \multicolumn{2}{c}{\textbf{Time}}            & \textbf{Cells}             \\\hline
Cyclic--3        & 9    & 3136               & 381              & 6     & 20 + 245 =&265               & 21                        \\ 
Cyclic--4        & 16   & $>1000s$           & ?                & 6     & 64 + 5813 =&5877             & 621                       \\
2                & 7    & 2249               & 895              & 14    & 22 + 1845 =&1867             & 579                       \\ 
4                & 6    & 3225               & 421              & 11    & 24 + 19738 =&19762           & 1481                      \\
6                & 4    & 363                & 41               & 5     & 20 + 918 =&938               & 89                        \\ 
7                & 8    & 3667               & 895              & 22    & 26 + 6537 =&6563             & 1211                      \\ 
8                & 6    & 3216               & 365              & 5     & 21 + 174 =&195               & 51                        \\ 
13               & 9    & 14342              & 4949             & 4     & 18 + 220 =&238               & 81                        \\
14               & 11   & 334860             & 27551            & 9     & 21 + 971 =&992               & 423                       \\\hline
\end{tabular}}
}
\documentclass{llncs}
\usepackage{url}
\usepackage{amsmath}
\usepackage{verbatim}
\usepackage{hyperref}
\usepackage{amssymb}
\usepackage[show]{ed}
\begin{document}
\title{Speeding up Cylindrical Algebraic Decomposition by Gr\"obner Bases}
\author{D.J.~Wilson, R.J.~Bradford \& J.H.~Davenport}
\institute{
Department of Computer Science, University of Bath\\
Bath BA2 7AY, U.K.\\ {\tt \{D.J.Wilson, R.J.Bradford, J.H.Davenport\}@bath.ac.uk}}
\maketitle
\begin{abstract}
Gr\"obner Bases \cite{Buchberger1970} and Cylindrical Algebraic Decomp\-osition \cite{Collins1975,Chenetal2009d} are generally thought of as two, rather different, methods of looking at systems of equations and, in the case of Cylindrical Algebraic Decomposition, inequalities.  
However, even for a mixed system of equalities and inequalities, it is possible to apply Gr\"obner bases to the (conjoined) equalities before invoking CAD. We see that this is, quite often but not always, a beneficial preconditioning of the CAD problem. 

It is also possible to precondition the (conjoined) inequalities with respect to the equalities, and this can also be useful in many cases. 

\end{abstract}
The examples used in this paper are available in \cite{Wilson2012a}.
This work was partially supported by the U.K.'s EPSRC under grant number EP/J003247/1.
\section{Introduction}
Solving systems of equations, or equations and inequations ($\ne$)/inequalities ($>,<$) is an old subject. Deciding the truth of, or more generally eliminating quantifiers from, quantified Boolean combinations of such statements, is more recent \cite{Tarski1951}. We can distinguish many families of methods, even if we restrict attention to the real numbers, or possibly the complex numbers.
\begin{description}
\item[\Gr]The method of Gr\"obner bases. Here the input is a set $S=\{s_1,\ldots,s_k\}$  of polynomials in some polynomial ring $k[x_1,\ldots,x_n]$ equipped with a total order\footnote{We have concentrated on purely lexicographical orders, since these seem to be the most useful to us.} $\prec$ on the monomials, and the output is a set $G=\{p_1,\ldots,p_l\}$ which is equivalent, in the sense that it generates the same ideal, i.e. $(G)=(S)$, and is simpler, or ``surprise-free'', in that the leading monomial with respect to $\prec$ (denoted $\lm_\prec$)  behaviour is explicit, $(\lm_\prec(G))=(\lm_\prec((G)))$. Then the solutions of $G$ are those of $S$, i.e. 
\begin{equation}\def\x{{\bf x}}
\left\{\x: p_1(\x)=0\land p_2(\x)=0\land\cdots\land p_l(\x)=0\right\}.
\end{equation}
\item[\Tr]The method of triangular decomposition via regular chains \cite{Aubryetal1999,MorenoMaza2005}. Here the output is a set of regular chains of polynomials \begin{equation}\{(p_{1,1},p_{1,2},\ldots),(p_{2,1},p_{2,2},\ldots),\ldots\},\end{equation} and the solution is the union of the set of regular zeros of these regular chains:
\begin{equation}\label{eq:RC}\def\x{{\bf x}}
\left\{\x: p_{1,1}(\x)= p_{1,2}(\x)=\cdots=0\land\left(\prod_i\init(p_{1,i})\right)(\x)\ne0\right\}\cup\cdots .
\end{equation}
\item[\Col]The method of Cylindrical (semi-)Algebraic Decomposition for real closed fields, computed via repeated projection to $\R^1$ and repeated lifting  \cite[and many improvements]{Collins1975}.
\item[\ECol]The previous case restricted to equality.
\item[\MMR]The method of  Cylindrical (semi-)Algebraic Decomposition for real closed fields via triangular decomposition \cite{Chenetal2009d}.
\item[\MMC]The method of  Cylindrical  Decomposition over the complexes via triangular decomposition, which was introduced in \cite{Chenetal2009d} as a stepping-stone to the previous method, but which probably has independent interest.
\item[\QECol]Quantifier Elimination by partial (i.e. taking account of the Boolean structure and quantifier structure) Cylindrical Algebraic Decomposition \cite{CollinsHong1991}.
\item[Others]such as Weispfenning's Virtual Term Substitution \cite[is a readable introduction]{Brown2005a}, or Tarski's original method \cite{Tarski1951}.
\item[Conversely]instead of asking for solutions $\bf x$ to $\exists {\bf x} f_1({\bf x})\ge0\land\cdots$, we may use a Positivstellensatz to show that no such ${\bf x}$ exist, as in \cite{Platzeretal2009}. We do not discuss this direction further here.
\end{description}
It should be noted that both \Col{} and \MMR/\MMC{} (but not \QECol) have the drawback that the Cylindrical Algebraic Decomposition produces decompositions for, not only the question posed, e.g. $\forall y\exists z p(x,y,z)=0\land q(x,y,z)=0\land r(x,y,z)>0$, but also {\it all\/} other questions involving the same polynomials, provided the quantifiers are over variables in the same order, e.g. $\exists y \forall z p(x,y,z)<0\lor(q(x,y,z>0\land r(x,y,z)=0)$. 
\par
This paper asks the question: ``can these methods usefully be combined?''  The combinations we are thinking about are those of conjunction: Can the fact that $B$ is in the context of $a_1=0\land\cdots\land a_k=0\land B$ be used to simplify $B$? In particular, we look at the use of Gr\"obner base methods to simplify the equalities in the conjunction {\it and\/} to simplify the inequalities in the light of the equalities.
\par
{\bf Technical Note:} all computations (\Gr, \MMR{} and \MMC) were performed in Maple 16$\beta$ on a 3.1GHz Intel processor, except for the \Col{}, \ECol{} and \QECol{} ones, which were performed on a 2.83GHz Intel processor with QEPCAD B version 1.65 \cite{Brown2003}. Times for a hybrid calculation, e.g. \Gr/\Col, are either quoted as the total time or a decomposition $a+b=c$ where $a$ is the time (in milliseconds) for \Gr, $b$ for \Col, and $c$ is the sum. We have run QEPCAD 
in three modes:
\begin{enumerate}
\item on the problem as given in \cite{BuchbergerHong1991}, implementing \QECol;
\item as above but with the \verb+full-cad+ option to ignore the Boolean structure of the expression;
\item with no quantifiers stated, and the \verb+full-cad+ option, implementing \Col.
\end{enumerate}
\section{Examples in this paper}
\subsection{\cite{BuchbergerHong1991}}
This paper has a variety of examples for \QECol, all of a form to which \Gr{} is applicable.
\subsection{\cite{Chenetal2009d}}
This paper has a variety of examples for \MMR. We chose some of those  to which \Gr{} is applicable.

\subsection{Two Spheres and A Cylinder}

Let the following be spheres in $\R^3$:
\begin{eqnarray*}
S_1: &\quad (x-1)^2 + y^2 + z^2 - 3; \\
S_2: &\quad (x+1)^2 + y^2 + z^2 - 3; \\
S_3: &\quad (x-1)^2 + \left(y-\frac{1}{2}\right)^2 + z^2 - 3; \\
S_4: &\quad (x+1)^2 + \left(y+\frac{2}{3}\right)^2 + \left(z + \frac{3}{4}\right)^2 - 3.
\end{eqnarray*}
Denote the infinite cylinder centred on the $z$-axis with radius 1 by $C$, so that the equation defining the cylinder is:
\begin{equation*}
C: \quad x^2 + y^2 - 1.
\end{equation*}

Now we investigate intersecting pairs of spheres (roughly increasing in CAD `difficulty') under conditions based on the cylinder. We assume the spheres' equation will always be required to equal 0 but make no assumptions on the condition on the cylinder. That is, we wish to solve the problem:
\begin{equation}
 S_i = 0 \ \wedge \ S_{i+1} = 0 \ \wedge \ C \ast 0 \qquad \ast \in \{=,\neq,<,>,\leq,\geq\}, i=1,2,3.
\end{equation}

We use the underlying variable ordering\footnote{This is the QEPCAD notation, meaning that we will project from $(z, y, x)$--space to $(z, y)$--space to $(z)$--space. We therefore end up with polynomials in $z$ alone, so this is equivalent to a purely lexicographical Gr\"obner base with $z\prec y\prec x$, i.e. {\tt plex([z,y,x])} in Maple: \GrC{} is used to indicate Gr\"obner bases with this (compatible) ordering. The CAD package in Maple \cite{Chenetal2009d} requires {\tt PolynomialRing([x,y,z])} to achieve the same effect as QEPCAD's $(z,y,x)$. \GrR{} denotes the reverse {\tt plex} order.} $(z,y,x)$.
\section{Prior Art}
Needless to say, we are not the first to have had this idea.
\subsection{Buchberger--Hong}
\cite{BuchbergerHong1991} considers the case of \Gr{} (\cite{Bogeetal1985} re-implemented in C) applied to \Col{} (an early version of \cite{CollinsHong1991} re-implemented in C), i.e., rather than computing a CAD for the zeros of a system of equations $E$ (i.e. $e_1=0\land e_2=0\land\cdots$) and inequalities $F$, compute it for $G$, a (purely lexicographical) Gr\"obner base for $E$, and $F$. They generally found a very substantial speed-up in the total computation time, e.g. ``Solotareff A''\footnote{There are various problems labelled ``Solotareff'': for a description of this class see \cite{Wilson2012a} and the links therein.}
\begin{eqnarray}
\exists x\exists y&&
3x^2-2x-a=x^3-x^2-ax-2b+a-2=\label{S1}\\&&
3y^2-2y-a=y^3-y^2-ay-2b+a-2=0\land\label{S2}\\&&
4a\in[1,7]\land 4b\in[-3,3]\land x\in[-1,0]\land y\in[0,1]\label{S3}
\end{eqnarray}
(with the variable ordering
$(b, a, x, y)$)
 took them 11500 ms for  \QECol{}, but 717 for \Gr{}, and 117 for \QECol{} applied to the result, a total of 834 ms, or a 13-fold speed-up. ``Solotareff B'' is the same problem but with $(a,b,x,y)$ as the variable ordering, and here the \QECol{} time was again greatly reduced, but the \Gr{} time was excessive. Of course, there have been substantial improvements in the implementation of all these algorithms since \cite{BuchbergerHong1991} was published, and Table \ref{Tab:BH-GB} shows that the \Gr{} time is now less than $1/3$ of the \QECol{} time. We choose rather to focus on the number of cells generated, which is closely connected to the \ECol{} time, and also affects the time taken to make use of the output. The cell counts are shown in Table \ref{Tab:Sol}.
\begin{table}[h]
\caption{Cell counts for Solotareff}\label{Tab:Sol}
\centerline{
\begin{tabular}{|ll|rr|rr|}\hline
&&\multicolumn{2}{c|}{\hfil Ordering A\hfil}&\multicolumn{2}{c|}{\hfil Ordering B\hfil }\\
&&\Col&\Gr/\Col&\Col&\Gr/\Col\\\hline
\hbox{(\ref{S1}--\ref{S3})}&Partial&153&63&375&41\\
&Full&349&625&1063&237\\\hline
\hbox{(\ref{S1}--\ref{S2})}&Partial&29&15&97&17\\
&Full&29&33&97&17\\\hline
\end{tabular}
}
\end{table}
\par More reruns of \cite{BuchbergerHong1991} are given in Table \ref{Tab:BH-GB}.
We see that, with today's technology, the conclusion of \cite{BuchbergerHong1991}, viz. that \Gr{} generally improves \QECol{} for the class of problems to which it is applicable, is still generally valid, but the details differ: notably the Gr\"obner base time is generally insignificant today.
\begin{table}[h]
\caption{\cite{BuchbergerHong1991} with today's technology}\label{Tab:BH-GB}
\tableA
\end{table}
\begin{table}
\caption{\cite{BuchbergerHong1991} Examples for full CADs}\label{BH-CAD}
\tableAb\hfil\break
\centerline{* indicates that the linear inequalities have been omitted in this version.}
\end{table}
\par
There is one point which is not explicit in \cite{BuchbergerHong1991}. As the computation of Gr\"obner bases in one variable is just equivalent to Euclid's algorithm, i.e. Gaussian elimination in Sylvester's matrix,  Gr\"obner base computations which are not genuinely multi-variate do not affect the set of resultants etc. generated in \Col, and hence are of limited use in the projection phase. They might still reduce the work done in the lifting phase, of course.
\par Table \ref{BH-CAD} re-runs the examples of \cite{BuchbergerHong1991}, but asking for complete cylindrical algebraic decompositions, and hence we can compare \Col{} with \MMR{} legitimately. Given that the algorithms are fundamentally different, the similarities in cell counts are striking. The differences in cell counts (where present) reflect differences in the cylindrical algebraic decompositions for the same input problem.
\subsection{Phisanbut}\label{sec:NP}
Phisanbut \cite{Phisanbut2011a}, considering branch cuts in the complex plane, observed that $g=0 \land f>0$ could be reduced to $g=0 \land \prem(f,g)>0$ under suitable conditions, where $\prem$ denotes the pseudo-remainder operation.  More precisely, if $f$ and $g$ are regarded as polynomials in the main variable $x$, of degrees $d$ and $e$ respectively, then $\prem(f,g)=\rem(c^{d-e+1}f,g)$, where $c$ is the leading coefficients of $g$. When $g=0$ and $c>0$, or when $d-e+1$ is even, $\prem(f,g)$ has the same sign as $f$. Unfortunately $c$ might have variable sign, and $d-e+1$ might be odd, so define $\pprecond(f,g)=\rem(c^{(d-e+1)^*}f,g)$, where $n^*$ is $n$ if $n$ is even and $n+1$ if $n$ is odd. Maple also defines $\sprem(f,g)=\rem(c^{m}f,g)$, where $m$ is the smallest integer such that the division is exact, and by analogy we have $\sprecond(f,g)=\rem(c^{m^*}f,g)$. Note that $\sprecond(f,g)=\pprecond(f,g)$ or a strict divisor of it, i.e. $\sprecond$ is never worse. She generally, but not always, saw \cite[Tables 8.13, 8.14]{Phisanbut2011a} a significant decrease in the number of cells, and the time taken to compute $\sprecond$ was minimal.
\section{Further developments}
\subsection{\Gr{} with \MMR}
It would seem natural to apply \Gr{} to \MMR, as \cite{BuchbergerHong1991} did to \QECol. The results are in Table \ref{BH-CAD}, and show a speed-up in all instances except the Collision problems. We also note the substantial speed advantage enjoyed by \Col, and this is a subject for further study.
\subsection{\Gr{} with \MMC}
We can also mix \Gr{} with \MMC, and these results are shown in Table \ref{TableB}, which also compares \MMC{} with \MMR. 
\begin{table}
\caption{Timings for \cite{BuchbergerHong1991} Examples: \MMR/\MMC}\label{Tab:BH-tr}\label{TableB}
\tableB\hfil\\
\centerline{``Ratio'' = (\MMR$-$\MMC)/\MMC, i.e. the relative cost of {\tt MakeSemiAlgebraic}.}
\end{table}
\MMR{} involves doing \MMC{} first, and then running the \verb+MakeSemiAlgebraic+ algorithm from \cite{Chenetal2009d}.  For these examples, the \verb+MakeSemiAlgebraic+ step is the most expensive initially, but  often not after we apply \Gr.
\subsection{\Gr{} with inequalities in \MMR}
Having reduced the equalities to a Gr\"obner base $G$, it is now possible to reduce the inequalities by $G$, since adding/subtracting a multiple of an element of $G$ is adding/subtracting 0. We can reduce with respect to the main variable, denoted \Gr/$\rightarrow^G_x$, with respect to secondary variables, denoted \Gr/$\rightarrow^G_y$ , or with respect to all variables (Maple's {\tt NormalForm}), denoted \Gr/\Greduce.
\begin{table}
\caption{Examples from \cite{Chenetal2009d}}\label{TableC}
\tableC
\end{table}
If we compare tables \ref{TableD} and \ref{TableE} we see that the number of cells produced is the same across the two methods.
\begin{table}
\caption{Spheres and Cylinders: \MMR}\label{TableD}
\tableD
\end{table}
\begin{table}
\caption{Spheres and Cylinders: \Col}\label{TableE}
\tableE
\end{table}
\section{Choice of Method}
Suppose we are given a problem, which we may formulate as 
\begin{equation}\label{eq:gen}
\hbox{\rm quantified variables } e_1=0\land\cdots\land e_k=0\land B(f_1,\ldots,f_l),
\end{equation}
where $B$ is a Boolean combination of conditions $=0,\ne0,<0$ etc. on some polynomials $f_j$, then we may be able, by applying Gr\"obner techniques to the $e_j$ , producing $e_j^{(i)}$, and then reducing the $f_j$, to produce various alternative formulations
$$
\hbox{\rm quantified variables } e^{(i)}_1=0\land\cdots\land e^{(i)}_{k^{(i)}}=0\land B(f^{(i)}_1,\ldots,f^{(i)}_l),
\eqno (\ref{eq:gen}^{(i)})
$$
and each of these may have several variable orderings compatible with the constraints implied by the quantification (if any). Which should we choose? Of course, in the presence of arbitrary parallelism, we can start them all, and accept the first to finish, but we may wish to be less extravagant.
\par
In the contexts of \QECol{} (strictly speaking, the REDLOG implementation), and where the only choice was in the variable order, this question was considered by \cite{Dolzmannetal2004a}. Retrospectively, there are two measures for the difficulty of a CAD computation: the time taken and the number of cells produced. For a given \QECol{} problem, they observed that two are usually correlated for different formulations, and we observe the same here for \MMR{} --- see our tables. However, we would like a measure that could be calculated in advance, rather than retrospectively.
\par
The processes of \cite{Collins1975,CollinsHong1991} starts with a set $A_n$ of polynomials in $n$ (ordered) variables $x_1,\ldots,x_n$, and
\begin{enumerate}
\item repeatedly project $A_i$ into $A_{i-1}$ in one fewer variable, until $A_1$ has only one variable,
\item[*](denote the set $\{A_n,\ldots,A_1\}$ by $A(x_1,\ldots,x_n)$)
\item isolate the roots of these polynomials to get a decomposition of $\R^1$,
\item repeatedly lift the decomposition until we get a (partial for \cite{CollinsHong1991}) cylindrical algebraic decomposition of $\R^n$.
\end{enumerate}
The third step is, both theoretically and practically, by far the most expensive. Hence the question arises: what can we measure at the end of step 1, i.e. depending on $A$ only, which is well-correlated with the final cost? Three things come to mind.
\begin{description}
\item[$\card(A(x_1,\ldots,x_n))$]$=\sum_{i=1}^n |A_i|$.
\item[$\td(A(x_1,\ldots,x_n))$]$=\sum_{i=1}^n\sum_{p_{i,j}\in A_i}\td(p_{i,j})$ where $\td$ denotes total degree. 
\item[$\sotd(A(x_1,\ldots,x_n))$]$=\sum_{i=1}^n\sum_{p_{i,j}\in A_i}\sum_{\hbox{monomials $m$ of }p_{i,j}}\td(m)$.
\end{description}
\cite{Dolzmannetal2004a} discard $\td$, observing that $\td$ and $\sotd$ are highly correlated and $\sotd$ ``has the advantage of favouring sparse polynomials''. They then observe that $\sotd(A(x_1,\ldots,x_n))$ is significantly more correlated with the retrospective measures for any given problem than $\card$. This gives a first algorithm for deciding how to project: for all admissible (i.e. compatible with the quantifier structure, if any) permutations $\pi$ of $(x_1,\ldots,x_n$), compute $A(x_{\pi(1)},\ldots,x_{\pi(n)})$, and choose the one with the least $\sotd$ value. The drawback of this is that it requires potentially $(n-1)n!$ projection operations. They show that (at least on their examples) this always produces a good projection order, and frequently the optimal.
\begin{table}
\caption{Spheres and Cylinders: \MMR{} --- choice of orderings}\label{TableFa}
\tableFa\break
\end{table}
\begin{table}
\caption{\cite{BuchbergerHong1991}: effect of orderings \GrC{} versus \GrR}\label{TableGa}
\tableGa\hfil\break
\centerline{We note that \GrR{} is definitely worse than \GrC.}
\end{table}
\par
They therefore propose a {\it greedy algorithm}, where for all permissible choices of the first variable to be projected, we compute $\sotd(A_{n-1})$, and choose the variable which gives the least value. Having fixed this as the first variable to project, for all permissible choices of the second variable to be projected, we compute $\sotd(A_{n-2})$, and choose the variable which gives the least value, and so on. Hence, assuming all projection orders are possible, the {\bf number} of projections done is $n+(n-1)+\cdots=O(n^2)$ rather than $n!$. It is currently an open question whether the {\bf cost} of projections behaves similarly.
\par
We proposed taking this idea still further, and suggested that, for several different formulations $A_n,B_n,\ldots$ of a problem, we should compute $\sotd(A_n)$, $\sotd(B_n),\ldots$ and take the formulation that yields the lowest \sotd.
We observed, however, that neither \td{} nor \sotd{} are good predictors in Table \ref{TableGb}, despite seeming useful in Table \ref{TableFb}.
\begin{table}
\caption{Spheres and Cylinders: \MMR{} ---degrees}\label{TableFb}
\tableFb\break
\centerline{`degrees' is \td$(A_n)$/\sotd$(A_n)$.}
\end{table}
\begin{table}
\caption{\cite{BuchbergerHong1991}: degrees}\label{TableGb}
\tableGb
\end{table}
\section{The metric {\tt TNoI}}

When we apply Gr\"obner techniques to a set of equations (either by calculating a basis or a normal form) we are, in some sense, trying to simplify the set of equations. In a zero-dimensional ideal, as shown in the Gianni-Kalkbrener Theorem \cite{Gianni1989,Kalkbrener1989}, a purely lexicographic Gr\"obner basis has a very distinct, triangular structure.

With this in mind we thought it may be of some use to consider the number of variables present in a certain problem and so defined the following quantity, {\tt TNoI}, which stand for ``Total Number of Indeterminates'':
\begin{equation}
{\tt TNoI}(F) = \sum_{f \in F} {\tt NoI}(f),
\end{equation}
where ${\tt NoI}(f)$ is the number of indeterminates present in a polynomial $f$. 
\subsection{{\tt TNoI} data}
The results of calculating such a quantity are given in Table 8, Table 9 and Table 10, showing a promising correlation to whether our preconditioning (with compatible ordering) is beneficial or not. In particular we note the following points:
\begin{itemize}
  \item In every example where preconditioning reduces {\tt TNoI} (15 cases) there is a significant reduction in timing (a decrease factor ranging from 4.20 to 757.26) and number of cells produced (a decrease factor ranging from 1.98 to 65.13).
  \item When preconditioning increases {\tt TNoI} (7 cases) then generally there is an increase in time (an increase factor ranging from 1.79 to 6.13) and the number of cells created (an increase factor ranging from 1.35 to 3.52) or the problem remains infeasible. There is one `false positive' result (\cite[Example 2]{Chenetal2009d}) where there is an increase in {\tt TNoI} but a slight improvement in the time (a decrease factor of 1.20) and cells produced (a decrease factor of 1.55).
  \item {\tt TNoI} alone does not measure the abstract difficulty of the calculations: Intersection A has a higher {\tt TNoI} than Ellipse A yet the latter takes 25 times longer and produces over 50 times as many cells. We have only shown how to use it to compare variants of the same problem.
\end{itemize}

As mentioned above, calculating {\tt TNoI} alone is not of a huge use, and even considering the difference or ratio does little to predict the degree of improvement to expect. However, if we take the logarithm of the ratio (equivalently the difference of the logarithms) of {\tt TNoI} and compare to the time or number of cells we get some interesting results. 

Plotting these quantities against each other certainly suggested there was a positive correlation. Recall that the sample correlation coefficient is defined as
\begin{equation}
  r_{X,Y} = \frac{\sum_{i=1}^n (X_i - \overline{X})(Y_i - \overline{Y})}{\sqrt{\sum_{i=1}^n (X_i - \overline{X})^2}\sqrt{\sum_{i=1}^n (Y_i - \overline{Y})^2}}
\end{equation}
and is a number between -1 and 1 that indicates how correlated data is. A sample coefficient of 1 indicates perfect positive correlation and a coefficient of -1 indicates perfect negative correlation. Although we are only working with a small bank of data (22 examples) and partially incomplete data (timings of $>1000$s were replaced by 10000 seconds and unknown cell numbers were replaced by 100000 to allow for coefficient calculation) there were still promising results. 

Let $S$ be the polynomial input, $\mathcal{D}_S$ its corresponding CAD, $t_S$ the time taken to calculate $\mathcal{D}_S$ and $c_S$ the number of cells in $\mathcal{D}_S$. Let $G$ be the Gr\"obner basis calculated with respect to the compatible ordering and define $\mathcal{D}_G$, $t_G$ and $c_G$ in a similar fashion. With the data set we obtained the sample correlation coefficients were as follows:
\begin{itemize}
  \item comparing $\log({\tt TNoI}(S)) - \log({\tt TNoI}(G))$ with $\log(t_S)-\log(t_G)$ gives a sample coefficient $r=0.821$ which indicates strong correlation (for our limited sample set).
  \item comparing $\log({\tt TNoI}(S)) - \log({\tt TNoI}(G))$ with $\log(c_S)-\log(c_G)$ gives a sample coefficient $r=0.829$ which again indicate a strong correlation (for our limited sample set). 
\end{itemize}

Of course correlation does not imply causation, especially with a relatively small data set, so let us look more deeply at what {\tt TNoI} is measuring. 

\begin{table}
\caption{TNoI for Spheres}
\tableH
\end{table}
\begin{table}
\caption{TNoI for \cite{BuchbergerHong1991}}
\tableI
\end{table}
\begin{table}
\caption{TNoI for \cite{Chenetal2009d}}
\tableJ
\end{table}

\subsection{What is {\tt TNoI} measuring?}

Consider what causes {\tt TNoI} to decrease. Let $S$ be a set of polynomials in variables $x_1,\ldots,x_n$ ordered $x_1 < x_2 < \cdots < x_n$. The following are three possible reasons for a decrease in {\tt TNoI}:
\begin{enumerate}
  \item The number of polynomials in a specific set of variables, $\{ x_{i_1},\ldots,x_{i_l}\}$, is decreased. If $x_k$ is the most important variable then reducing the number of these polynomials will simplify the decomposition in the $(x_1,\ldots,x_k)$-plane. This will simplify the overall CAD, reducing the number of cells produced and hence the time taken to calculate the decomposition.
  \item At least one variable is eliminated from a polynomial. If the variable $x_k$ is eliminated from a polynomial $p$ then the decomposition based around $p$ will be greatly simplified. This will again simplify the overall CAD, reducing the number of cells produced and hence the time taken to calculate the decomposition.
  \item A polynomial in a large number of variables, say $k$, is replaced by $j$ polynomials each with $n_i$ variables such that $\sum n_i < k$. Intuitively this would increase the {\it number\/} of discriminants and resultants calculated, be it in the projection phase of \Col{} or in \MMC, but the results appear in lower levels of the projection tree, and this effect is more potent than the apparent increase in the number of discriminants and resultants. We have yet to build a good model of this, though.
\end{enumerate}
Obviously, in general, a combination of these factors will be the reason for the decrease in {\tt TNoI}. Also, there may be opposing increases in {\tt TNoI}, which presumably explains why the `false positive' of \cite[Example 2]{Chenetal2009d}  shows an increase in {\tt TNoI} but an improvement in the CAD efficiency.
\section{Conclusions}
\begin{itemize}
\item For both \Col{} and \MMR{} and \MMC, pre-conditioning the equations (where applicable) by means of a Gr\"obner calculation is often well worth doing.
\item Gr\"obner reduction of inequalities with respect to equalities has never, in our examples, made things worse.
\item {\it A priori\/}, it can be quite difficult to see which combinations of  Gr\"obner base and Gr\"obner reduction will be best, but the Gr\"obner side is generally cheap\footnote{This is a significant change from \cite{BuchbergerHong1991}, who had examples where the Gr\"obner calculations was much more expensive than the Cylindrical Algebraic Decomposition.}.
\item We therefore have multiple equivalent formulations of a given problem. We have investigated the metrics of \cite{Dolzmannetal2004a}, but have concluded that, at the level of choice of formulation, {\tt TNoI} is a better predictor. It does not help for predicting the best ordering of variables, for which \cite{Dolzmannetal2004a} or the Brown heuristic \cite{Brown2004} are appropriate. Phisanbut \cite[Chapter 8]{Phisanbut2011a} found the Brown heuristic sufficiently good, and simpler to compute.
\item In Section \ref{sec:NP} we saw how $g=0 \land f>0$ could be reduced to $g=0 \land \sprecond(f,g)>0$. In principle, given $s_1=0\land\cdots\land s_k=0\land f>0$, after computing  a Gr\"obner base $G$  for the $s_i$, we could attempt a more general reduction of $f$ by $G$.  Pure {\tt NormalForm} reduction has proved useful (Tables \ref{TableD}, \ref{TableE}), but we do not have enough good examples to measure the utility of a more general pseudoremainder-like reduction.
\end{itemize} 

\end{document}